\documentclass[12pt,a4paper]{article}
\usepackage{localeng}

\newcommand{\cnd}{\mskip 0.5mu{|}\mskip 0.5mu}
\DeclareMathOperator{\KS}{\mathrm{C}\mskip 0.1mu}
\DeclareMathOperator{\KP}{\mathrm{K}\mskip 0.1mu}

\let\ge=\geqslant
\newtheorem{theorem}{Theorem}

\begin{document}
\init

\title{Kolmogorov complexity as a combinatorial tool}
\author{Alexander Shen\thanks{LIRMM, Univ Montpellier, CNRS, Montpellier, France \texttt{alexander.shen@lirmm.fr}, \texttt{sasha.shen@gmail.com}, \protect\url{https://lirmm.fr/~ashen}, ORCID 0000-0001-8605-7734. Supported by 
ANR grant ANR-21-CE48-0023 FLITTLA.}}

\maketitle
\begin{abstract}
Kolmogorov complexity is often used as a convenient language for counting and/or probabilistic existence proofs. However, there are some applications where Kolmogorov complexity is used in a more subtle way. We provide one (somehow) surprising example where an existence of a winning strategy in a natural combinatorial game is proven (and no direct proof is known).

%\keywords{Kolmogorov complexity \and Combinatorial games \and Experts aggregation}
\end{abstract}

\section{Introduction}

It is well known that Kolmogorov complexity is a useful tool for proving combinatorial statements; the entire Chapter 6 (\emph{The Incompressibility Method}) of the classical Li--Vitányi textbook~\cite{lv} is devoted to applications of this type --- too numerous to mention them here. A typical application of the incompressibility method can be described as follows. We want to prove the existence of an object  $x$ in some class $X$ such that $x$ has some property $P(x)$. For that we show that every object $x\in X$ \emph{not} having this property is compressible (has small Kolmogorov complexity) while most of the objects in $X$ are incompressible. 

A toy example: most undirected graphs with $n$ vertices have complexity about $n^2/2$ (for each pair of vertices we need to specify whether there is an edge between them), but graphs that have clique or independent set $S$ of large size $k$ have shorter descriptions.  Indeed, we do not need to consider individually all the pairs in $S$ (about $k^2/2$ bits economy), it is enough to specify $S$ itself ($k$ vertices, $\log n$ bits per vertex), so if $k \gg \log n$ we get a shorter description (since $k^2\gg k\log n$) . Therefore, $n$-vertex graphs of maximal complexity have no cliques or independent sets of size $O(\log n)$ --- so we have proven the existence of graphs without large cliques and independent sets.

This argument is essentially a counting argument showing that the number of graphs that have a large clique or independent set is much smaller than the total number of graphs with $n$ vertices. Many other applications of Kolmogorov complexity can be translated to a counting argument in the same way (which does not mean, of course, that Kolmogorov complexity is not needed: a useful language is very important).   More generally, we may use probability with respect to some non-uniform distribution instead of counting to estimate Kolmogorov complexity. For example, to prove the law of large numbers we may note that if the frequency of ones in a bit string $x$ deviates from $1/2$, one can use skewed Bernoulli distribution and the relation between Kolmogorov complexity $\KS(x)$ and a priori probability of $x$ (or just arithmetic coding) to show that the string is compressible. In other proofs we use a distribution on the outputs of some probabilistic algorithm and show that with high probability the algorithm produces an object we are looking for (otherwise the sequence of random bits used by the algorithm is compressible).  Two proofs of this type are reproduced and discussed in~\cite{shen-language}: the compressibility proof of (a special case of the) Lovász local lemma and the ``tetris'' proof of the existence of a sequence that avoids forbidden factors.

Are there other applications of Kolmogorov complexity to combinatorics\footnote{\emph{Disclaimer}: we do not consider applications of Kolmogorov complexity to \emph{infinite} objects --- e.g., point-to-set principle in geometric measure theory or the notion of $\KP$-trivial sequence in recursion theory; see also other examples in~\cite{shen-language}.} that do not follow this scheme?  Can we reformulate these arguments without Kolmogorov complexity or at least find some other arguments that do not use complexity?

Here are some examples of this type:

\begin{itemize}

\item Levin's lemma from~\cite[Section 4.2]{dls} says that there exists a binary sequence such that all its sufficiently long factors are almost incompressible. This is the complexity version of a statement about forbidden factors and can be proven using Lovász local lemma (see~\cite{ru}). The original Levin's proof involves multiple uses of the Kolmogorov--Levin formula for complexity of pairs, so its translation into combinatorial language is difficult; however, the combinatorial argument given by Rosenfeld~\cite{rosenfeld} (see also Muchnik's argument in~\cite[p.~259]{suv}) can be considered (at some extent) as a combinatorial counterpart of Levin's argument in a much more general setting with much stronger bounds.

\item There exists a simple proof of Loomis--Whitney inequality using Kol\-mo\-go\-rov complexity (see~\cite[Section 8.8 and Chapter 10]{suv}); the complexity argument again uses the Kolmogorov--Levin formula for complexity of pairs several times, so its combinatorial translation is difficult. However, the proof can be reformulated easily in terms of Shannon entropy.

\item A combinatorial result saying that every multidimensional set can be covered by a small number of ``almost uniform'' subsets can be derived from the classification of tuples according to their ``complexity profile'' in a quite straightforward way, see~\cite[p.~333, Lemma and its proof]{suv}. It is unclear how this argument can be translated into a combinatorial or probabilistic language --- however, a combinatorial proof of a stronger version of this result exists~\cite{uniform}.

\end{itemize}

There is one more example of a complexity result where more advanced tools are used, and its combinatorial corollary for which a combinatorial proof is not known (though a weaker result can be proved directly). 

\section{New example}

This complexity statement deals with combinatorial rectangles that are simple with respect to each its element. A set $R\subset \mathbb{B}^n\times\mathbb{B}^n$ (pairs of $n$-bit strings) is a \emph{combinatorial rectangle} if $R=U\times V$ for some $U,V\subset \mathbb{B}^n$. For every finite set $A$ of constructive objects let 
\(
i(A)=\min_{x\in A} \KS(A\cnd x),
\)
where $\KS(A\cnd x)$ is the conditional Kolmogorov complexity of $A$ given $x$. Intuitively speaking, sets $A$ with small $i(A)$ are classes of ``good classifications'': any equivalence class $A$ for an equivalence relation of small complexity has small $i(A)$, since we can find $A$ given the relation and arbitrary element of $A$.

\begin{theorem}[Romashchenko, Zimand, lemma 4.6 in~\cite{rz} adapted to one rectangle]
\[
\KS(R)\ge \KS(R\cnd x) + \KS(R\cnd y) - O(\log n + i(R))
\]
for every combinatorial rectangle $R\subset \mathbb{B}^n\times\mathbb{B}^n$ and every its element $(x,y)\in R$.
\end{theorem}

The original motivation for this result was the communication complexity setting where $R$ is the combinatorial rectangle that corresponds to some transcript of the communication protocol for inputs $x$ and $y$. However, it is a general complexity statement that can be proven using the complexity version of artificial independence trick (``copy lemma'') used to prove non-Shannon inequalities (see, e.g.,~\cite[Section 10.13]{suv} for more details).

To get a combinatorial translation, we can use Muchnik's game interpretation of Kolmogorov complexity results described in~\cite{muchnik}. We do not describe the details, just formulate the result in terms of a winning strategy for a simple game.  The game field is $\mathbb{B}^n\times\mathbb{B}^n$; two players alternate. The first player provides a sequence of disjoint combinatorial rectangles in $\mathbb{B}^n\times\mathbb{B}^n$. After a new rectangle is chosen, the second player replies by labeling this rectangle as ``horizontal'' or ``vertical''. The games continues for $T=ab$ moves (where $a$ and $b$, as well as some factor $k$, are parameters of the game), and the second player wins if every horizontal line $\mathbb{B}^n\times\{y\}$ intersects at most $ka$ horizontal rectangles, and every vertical line $\{x\}\times \mathbb{B}^n$ intersects at most $kb$ vertical rectangles.

\begin{theorem}
For some polynomial $p(n)$, the second player has a winning strategy in this game for all $n,a,b$ and $k=p(n)$.
\end{theorem}

Alexander Kozachinskiy and Tomasz Steifer (personal communication) noted that this game has a interesting version in terms of opinions' aggregation. Imagine a group of people traveling in a train. At every stop the train conductor comes and ask passengers whether they want the heating to be on or off (till the next stop). Some people want it on, some other want it off, some do not care and are happy with both options. Every decision of the conductor makes some passengers unhappy (those who wanted the opposite option), and this is unavoidable if there are conflicting requests, but the conductor wants to minimize the \emph{maximal unhappiness} among passengers. (The unhappiness of a passenger is the number of her requests that were not fulfilled.) This can be difficult: for example, if two passengers have conflicting requests all the time (on every stop), then for $t$ stops at least one of them has unhappiness at least $t/2$ (for obvious reasons). However, the following result is true for some polynomial $p$:

\begin{theorem}[Kozachinskiy--Steifer]
If every pair of passengers has a conflict at most once, then the conductor can guarantee that the maximal unhappiness after $t$ stops does not exceed $\sqrt{t} \cdot p(\log n)$ where $n$ is a number of passengers.
\end{theorem}

This statement can be reduced to the rectangle game. Informally speaking, the game board consists of passenger pairs, and for each stop we consider a combinatorial rectangle
\[
 (\text{people who want the heating}) \times  (\text{people who  do not want the heating}),
\]
the requirement about conflicts guarantees that the rectangles are disjoint, and vertical/horizontal labeling of rectangles corresponds to the on/off actions.

Kozachinskiy and Steifer found a combinatorial proof of a weaker version of this result, with $t^{2/3}$ instead of $\sqrt{t}$, using some modification of Littlestone--Warmuth weighted majority algorithm. Their proof provides an explicit (and computationally simple) strategy for the conductor while the complexity argument gives only the existence proof without any complexity bound (except for a trivial one that corresponds to the exhaustive search).

\subsection*{Constants in the definition of complexity}

Finally, let us make some general remarks about complexity approach to combinatorics. The Kolmogorov complexity function is defined up to $O(1)$ additive terms: when we change the optimal programming language used in its definition, the numerical value of the complexity changes, but the changes are bounded by a constant that depends only on the two programming languages that we compare. This seems unavoidable, and people are accustomed to it. However, if we prove a combinatorial corollary that does not mention the complexity explicitly, we would like to be more explicit about the constant.

In the simplest applications of the incompressibility method this is not a problem: we construct some decompressor (that does not even need to be universal) and note that all the ``bad objects'' (e.g., the graphs with large clique or independent set) are compressible in this sense, while most graphs are incompressible (the later statement is valid for \emph{every} decompressor and does not involve unknown constants). However, in more advanced arguments, for example when we use the Kolmogorov--Levin formula for complexity or pairs, this simple trick does not work. In these arguments we also get an $O(\log n)$ term that often translates to polynomial factors in the combinatorial setting, and the constant in the $O(\log n)$ notation can be quite big. Fortunately, this constant does not depend on the choice of the programming language, but $O(1)$ terms do, so to get a specific bound directly from the argument we need to fix a universal programming language and actually write programs involved in the proofs using this language. This is boring and most probably gives unreasonable large constants. Maybe one can choose some special programming language to minimize the efforts (and/or constants)?
 
Let us note also that in some arguments (e.g., for the Loomis--Whitney argument, and for the arguments with randomized algorithms mentioned above) the constants in the definition of Kolmogorov complexity do not matter (since the construction used in the argument includes some asymptotic reasoning).

\textbf{Acknowledgments}. This paper was prepared for Special Session invited talk at Computability in Europe 2024 conference.
 The author is grateful to the section organizers for the invitation, to Alexander Kozachinskiy and Tomasz Steifer for the permission to include their unpublished results, and to the participants of the Kolmogorov seminar and LIRMM colleagues (especially Andrei Romashchenko and Matthieu Rosenfeld) for interesting discussions.


\begin{thebibliography}{9}

\bibitem{lv}
Ming Li, Paul Vitányi, \emph{An Introduction to Kolmogorov Complexity and Its Application}, 4th ed., Springer, 2019. 

\bibitem{dls}
B.~Durand, L.~Levin, A.~Shen, Complex Tilings, \emph{Journal of Symbolic Logic}, \textbf{73}(2), 593--613 (2008)

\bibitem{uniform}
 N.~Alon, I.~Newman, A.~Shen, G.~Tardos, N.~Vereshchagin, Partitioning multi-dimensional sets in a small number of ``uniform'' parts, \emph{European Journal of Combinatorics, \textbf{28}(1}, 134--144 (207), \url{https://www.sciencedirect.com/science/article/pii/S0195669805001204}, see also \url{https://eccc.weizmann.ac.il/report/2005/095/}

\bibitem{muchnik}
Andrej A.~Muchnik, Ilya Mezhirov, Alexander Shen, Nikolay Vereshchagin, \emph{Game interpretation of Kolmogorov complexity} (2010), \url{https://arxiv.org/pdf/1003.4712}

\bibitem{rz}
 A.~Romashchenko, M.~Zimand, An Operational Characterization of Mutual Information in Algorithmic Information Theory, \emph{Journal of the ACM}, \textbf{66}(5), 1--42 (2019), \url{https://dl.acm.org/doi/10.1145/3356867}
 
 \bibitem{rosenfeld}
 M.~Rosenfeld, \emph{Finding lower bounds on the growth and entropy of subshifts over countable groups},  \url{https://arxiv.org/pdf/2204.00394}
 
 \bibitem{ru}
 A.Yu.~Rumyantsev, M.~Ushakov, Forbidden strings, Kolmogorov Complexity and Almost Periodic Sequences, \emph{STACS 2006, 23rd Annual Symposium on Theoretical Aspects of Computer Science, Marseille, France, February 23--25, 2006, Proceedings} (Lecture Notes in Computer Science, \textbf{3884}), Springer, 2006, 396--407, \url{https://doi.org/10.1007/11672142}, see also~\url{https://arxiv.org/pdf/1009.4455}.
 
 \bibitem{shen-language}
 A.~Shen, Kolmogorov complexity as a language, \emph{Computer Science --- Theory and applications}, CSR 2011 Proceedings, Springer, 2011. (Lecture Notes in Computer Science, \textbf{6651}), 105--119, \url{https://doi.org/10.1007/978-3-642-20712-9_9}, see also \url{https://arxiv.org/pdf/1102.5418}
 
 \bibitem{shen-cie}
 A.~Shen, Compressibility and Probabilistic Proofs, \emph{Unveiling Dynamics and Complexity -- 13th Conference on Computability in Europe, CiE 2017, Turku, Finland, June 12--16, 2017, Proceedings} (Lecture Notes in Computer Science, \textbf{10307}), Springer, 1997, see also \url{https://arxiv.org/pdf/1703.03342}.
 
 \bibitem{suv}
 A.~Shen, V.A.~Uspensky, N.~Vereshchagin, \emph{Kolmogorov complexity and algorithmic randomness}, American Mathematical Society (Mathematical Surveys and Monographs, volume 220), \url{https://www.lirmm.fr/~ashen/kolmbook-eng-scan.pdf}

\end{thebibliography}
\end{document}